\documentclass[prd,12pt,aps,preprintnumbers,tightenlines,superscriptaddress,showpacs,floatfix,nofootinbib,amsmath,amssymb]{revtex4}
\usepackage{graphicx}
\usepackage{amssymb}

\newcommand{\dd}{{\mathrm{d}}}
\newcommand{\Tr}{{\mathrm{Tr}}}
\newcommand{\NP}{{\mathrm{NP}\!}}
\newcommand{\bbbone}{{\mathchoice {\rm 1\mskip-4mu l} {\rm 1\mskip-4mu l}{\rm 1\mskip-4.5mu l} {\rm 1\mskip-5mu l}}}
\newcommand{\cF}{{\mathcal F}}
\newcommand{\cP}{{\mathcal P}}
\newcommand{\cC}{{\mathcal C}}
\newcommand{\cS}{{\mathcal S}}
\newcommand{\cG}{{\mathcal G}}
\newcommand{\logo}{\\ \vskip -26mm \leftline{\includegraphics[scale=0.3,clip=false]{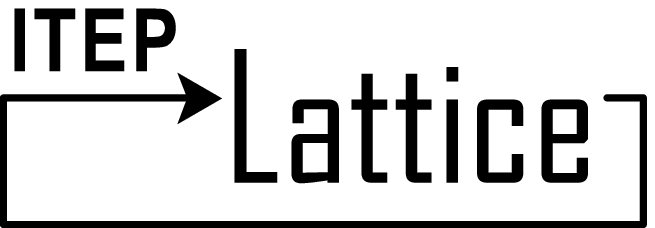}} \vskip 20mm}

\begin{document}

\preprint{ITEP-LAT/2008-02}

\title{Topological density fluctuations and gluon condensate \\
around confining string in Yang--Mills theory\logo
}

\author{M. N. Chernodub}
\affiliation{ITEP, B. Cheremushkinskaya 25, Moscow 117218, Russia}
\author{I. E. Kozlov}
\affiliation{ITEP, B. Cheremushkinskaya 25, Moscow 117218, Russia}
\affiliation{M.V. Lomonosov Moscow State University, Faculty of Physics, Moscow, 119992, Russia}

\begin{abstract}
We study the structure of the confining string in Yang-Mills theory using the method of
the field strength correlators. The method allows us to demonstrate that both the local
fluctuations of the topological charge and the gluon condensate are suppressed in the
vicinity of the string axis in agreement with results of lattice simulations.
\end{abstract}


\date{\today}

\maketitle

\section{Introduction}
The formation of the chromoelectric string between quarks and antiquarks is widely accepted as the reason
of quark confinement in QCD. The string around the static quarks is indeed found in numerical simulations
of Yang--Mills theory on the lattice~\cite{Bali:review}. On physical grounds the quark confinement
is understood as follows: the chromoelectric fluxes coming from the quarks and sinking into the
antiquarks are squeezed into the vortexlike (string) structures. Since the chromoelectric vortices
possess a non-zero tension, they provide a confining force between the quarks and antiquarks.
This force makes the quarks to be confined into colorless hadrons.

The string structure shows its imprint on distribution of various local observables. The lattice
simulations of Yang--Mills theory demonstrate squeezing of the chromoelectric fluxes spanned between
static quarks~\cite{Bali:1994de,ref:string:other}. The same numerical method shows the expected
enhancement of the chromoelectric energy density
at the string axis~\cite{ref:string:energy-action,Bissey}. In the contrary, the gluonic action density
(or, equivalently, the gluon condensate) is suppressed by the string in agreement with the sum rules
predictions~\cite{ref:sum-rules}.
The strong chromoelectric field of the confining string affects also various local condensates~\cite{Chernodub:2005gz}.

On general grounds one may expect that the string should make an influence on the topological charge
density as well. A quantitative characteristic of this effect may be provided by
the quantum expectation average of the {\it squared} topological charge density because the first
power of the topological density always vanishes due to the CP invariance of the pure Yang--Mills theory.
The expectation value of the squared topological charge density provides a measure of local
fluctuations of the topological charge and therefore hereafter we refer to this expectation value
as to ``the local susceptibility''. A comparison of the local topological susceptibility
calculated in close vicinity of the string and far away from the string gives a measure of
the impact of the chromoelectric strings on the topological properties of the vacuum.

The early numerical simulations -- which used a cooling procedure in order to push the quantum fields
towards a smooth classical limit -- have indeed revealed a suppression of the local fluctuations of the
topological charge density at the string axis~\cite{string:topcharge:first,string:topcharge}. The
subsequent study of the topological susceptibility in the uncooled vacuum confirmed the suppression
of the topological susceptibility at the string axis~\cite{ref:ourwork}. In addition it was observed that
as the string gets longer the suppression region widens in the transverse direction with respect to the
string axis~\cite{ref:ourwork}. Later, the widening in terms of the action density was precisely studied in
Ref.~\cite{ref:BGM}. The widening of the string shape -- as probed both by the topological charge density
and by the gluon condensate -- happens due to the quantum fluctuations of the string. The string fluctuates
transversely to its axis in such a way that the geometrical center of the string follows the Gaussian
distribution as it was predicted by L\"uscher, M\"unster and Weisz~\cite{Luscher:1980iy}.

In this paper we study the structure of the confining string using the analytical method of the field
strength correlators~\cite{ref:correlators}. This gauge-invariant method
is a powerful tool to calculate various nonperturbative features of the Yang--Mills vacuum (for reviews see
Refs.~\cite{ref:review,ref:review:two}) including the quantities
related to the topological charge density as well. Indeed, the topological susceptibility -- given by a
volume integral of the two-point correlator of topological densities -- can be calculated both on the
lattice with the use of numerical methods and in the continuum limit, analytically. In the latter case
the field correlator technique can be applied in rather straightforward manner~\cite{ref:susceptibility}.
The analytical and numerical results coincide with each other. Note that in the real QCD vacuum the
correlator of the two topological densities receives a substantial contribution from light quarks~\cite{ref:BL}.

It is worth noticing that the effects of the string fluctuations -- leading to widening of the string and
to the substantial smearing of the topological density and gluon condensate transverse to the string axis
-- are not taken into account in our study. We investigate the ``bare''  nonfluctuating string which possess
the finite width due to the squeezing properties of the confining vacuum.

\section{Method of field correlators}
The central object of the method is the element of the $su(N)$ algebra
\begin{equation}
\hat \cF_{\mu\nu}(x;x_0) = \Phi(x_0;x) {\hat F}_{\mu \nu}(x) \Phi(x;x_0)\,,
\label{eq:G}
\end{equation}
where $\hat F_{\mu\nu} = F_{\mu\nu}^a T^a$ is the non-Abelian field strength tensor
$F^a_{\mu\nu} = \partial_\mu A^a_\nu -  \partial_\nu A^a_\mu + g f^{abc} A^b_\mu A^c_\nu$, and
$T^a$, $a=1,\dots,N^2-1$ are the generators of the $SU(N)$ gauge group. We use the ``hat'' over
variables which take their values in the $su(N)$ algebra. In Eq.~(\ref{eq:G}) the nonlocal
quantity $\Phi(x;x_0) \equiv \Phi^\dagger (x_0,x)$ represents the Schwinger line,
\begin{equation}
\Phi(x;x_0) = \cP \exp\Bigl[ - i g \int\nolimits_x^{x_0} \hat A_\mu (z) \, d z_\mu \Bigr]\,.
\label{eq:Phi}
\end{equation}
Here the integration of the gauge field $\hat A_\mu = T^a A^a_\mu$ is taken
along the oriented path $\cC_{x,x_0}$ spanned between the points $x_0$ and $x$ of the
Euclidean space-time. In order to enforce the gauge covariance the exponent in Eq.~(\ref{eq:Phi})
is subjected to the path ordering $\cP$. Therefore under the $SU(N)$ gauge transformation
$\Omega$ the quantity~(\ref{eq:Phi}) transforms covariantly at points $x$ and $y$:
\[
\Phi(x;y) \to \Omega(x) \Phi(x;y) \Omega^\dagger(y)\,.
\]
Due to the presence of the Schwinger line the nonlocal object~(\ref{eq:G}) behaves
at the reference point $x_0$ as a local covariant quantity:
\[
\hat \cF_{\mu\nu}(x,x_0) \to \Omega(x_0) \hat \cF_{\mu\nu}(x,x_0) \Omega^\dagger(x_0)\,.
\]
Since the object $\hat \cF_{\mu\nu}(x,x_0)$ transforms in the adjoint representation of the gauge group
one can construct various gauge invariant quantities, among which the simplest two-point correlator
\begin{equation}
D_{\mu\nu\alpha\beta}(x_1,x_2;x_0) = \frac{g^2}{N} \Tr [\hat \cF_{\mu\nu}(x_1,x_0) \, \hat \cF_{\alpha\beta}(x_2;x_0)]\,,
\label{eq:D2}
\end{equation}
plays the most important r\^ole. The dependence of the correlator on the reference point $x_0$ can be omitted
if we choose all Schwinger paths to be segments of the straight line connecting the points $x_1$ and $x_2$.
After this procedure the correlator~(\ref{eq:D2}) becomes a function of the single variable $z = x_1 - x_2$.

The correlation function~(\ref{eq:D2}) can in general be represented as follows
\begin{eqnarray}
D_{\mu\nu\alpha\beta}(z) & = & \Bigl(\delta_{\mu\alpha} \delta_{\nu\beta}
- \delta_{\mu\beta} \delta_{\nu\alpha} \Bigr) \, D(z^2) \label{eq:g2D}
\\
& & + \frac{1}{2} \Bigl[
\frac{\partial}{\partial z_\mu} (z_\alpha \delta_{\nu\beta} - z_\beta \delta_{\nu\alpha}) -
\frac{\partial}{\partial z_\nu} (z_\alpha \delta_{\mu\beta} - z_\beta \delta_{\mu\alpha})
\Bigr]\, D_1(z^2)\,,
\nonumber
\end{eqnarray}
where $D$ and $D_1$ are the scalar structure functions. According to the lattice
simulations~\cite{ref:Adriano}, the structure functions can
be described by the ansatz
\begin{equation}
D_i(z^2)   = A_i \, e^{ - |z|/T_g} + \frac{b_i}{|z|^4} \, e^{ - |z|/\lambda} \,, \quad i = 0,1\,,
\label{eq:D:D1}
\end{equation}
where the correlation lengths $T_g$ and $\lambda$ as well as the prefactors $A_i$ and $b_i$, $i=0,1$ can be
determined from the lattice data, and $D_0 \equiv D$. The first terms in Eq.~(\ref{eq:D:D1}) correspond to a nonperturbative
contribution while the last terms contain the perturbative $1/|z|^4$ divergencies at short distances, $|z| \to 0$.

The simplest non-vanishing correlator~(\ref{eq:g2D}) determines a dominant contribution to various nonperturbative
observables~\cite{ref:review,ref:check}. According to the stochastic scenario all higher order connected correlators
are suppressed with respect to the leading Gaussian contribution, while all $2n$--point correlators can be
expressed via the bilocal correlator~(\ref{eq:g2D}).

\begin{figure}[htb]
\begin{center}
\begin{tabular}{cc}
\includegraphics[scale=1.3,clip=false]{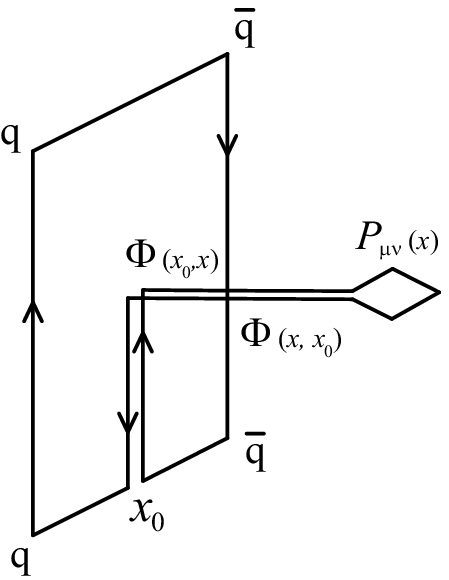} &
\hspace{5mm} \includegraphics[scale=1.3,clip=false]{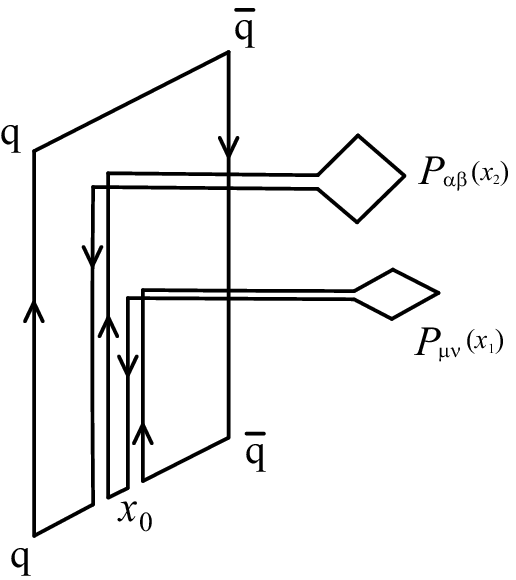} \\
\end{tabular}
\end{center}
\caption{The one-plaquette (left) and two-plaquette (right) probes.}
\label{fig:probes}
\end{figure}

In order to study the structure of the chromoelectric string we follow Refs.~\cite{ref:SK00} and
\cite{ref:SK:more}.
First, we consider the gauge--invariant quantity given by the quantum average
\begin{equation}
\cF^{\cC}_{\mu\nu}(x)
\equiv
\langle \hat \cF_{\mu \nu}(x) \rangle_{\cC}
\equiv
\frac{\langle {\mathrm{Tr}}\, \hat \cF_{\mu \nu}(x,x_0) \,
W(\cC_{x_0}) \rangle}{\langle { \mathrm{Tr}} \, W(\cC) \rangle} - 1\,,
\label{eq:one:probe:general}
\end{equation}
which is depicted schematically in Figure~\ref{fig:probes}(top). The Wilson loop $\cC_{x_0}$ is creating the
static meson made of the quark $q$ and the antiquark $\bar q$ separated by the distance $R$. The world--trajectory of the
pair -- which is usually chosen to be a rectangular $\cC = T \times R$ -- is open at the point $x_0$,
to which the two Schwinger lines, $\Phi(x_0,x)$ and $\Phi(x,x_0) \equiv \Phi^\dagger(x_0,x)$ are attached. The
Schwinger lines connect the Wilson loop with the elementary plaquette $P_{\mu\nu}(x)$ making the whole object
explicitly gauge--invariant. The form of the Schwinger lines can readily be seen in Figure~\ref{fig:probes}(top).
In the limit of vanishing plaquette size the whole construction (if normalized properly) gets equal
to Eq.~(\ref{eq:one:probe:general}).
The meaning of the quantity~(\ref{eq:one:probe:general}) is straightforward: it probes at the point $x$
the $\mu\nu$-component of the gluonic flux which is induced by the quark $q$ and the
antiquark $\bar q$ traveling along the closed path $\cC$. Below we study a permanently living static meson
corresponding to the limit $T \to \infty$.

The quantity~(\ref{eq:one:probe:general}) can be calculated in the Gaussian (bilocal) approximation and gives
the result~\cite{ref:SK00}:
\begin{equation}
\cF^\cC_{\mu \nu}(x) = - \int_S d \sigma_{\alpha \beta} (y) D_{\mu \nu \alpha \beta}(x - y)\,.
\label{eq:one:probe:flux}
\end{equation}
Hereafter next-to-leading corrections are not shown. The analytical formula~(\ref{eq:one:probe:flux}) --
supplemented by the ansatz~(\ref{eq:g2D}) and by the parameterization~(\ref{eq:D:D1}) -- indicates the squeezing of the
chromoelectric flux between the quarks into the tube of a finite width. Moreover, one can easily find that
the vacuum expectation value (v.e.v.) of the chromomagnetic field is zero for the static string. Thus, the authors
of Ref.~\cite{ref:SK00} have clearly demonstrated the formation of the confining chromoelectric string using the
approach of the field strength correlators.

\section{Gluon condensate around the string}
This paper is devoted to the generalization of the successful approach of Refs.~\cite{ref:SK00,ref:SK:more}
with respect to other quantities. The most interesting such quantity is the
topological susceptibility. However, let us first consider the imprint of the string on the gluon condensate
which is simpler observable compared to its topological counterpart.

In order to calculate the string shape in terms of the gluon condensate we use the connected correlator of
a special configuration, which includes two distinct plaquette probes:
\begin{eqnarray}
\cS^{\cC}_{\mu \nu \alpha \beta} & = &
\frac{g^2 N}{2 \pi^2} \frac{\langle {\mathrm{Tr}}\, \hat \cF_{\mu \nu}(x_1,x_0) \, \hat \cF_{\mu \nu}(x_2,x_0) \,
W(\cC_{x_0}) \rangle}{\langle { \mathrm{Tr}} \, W(\cC) \rangle}\,,
\nonumber \\
& \equiv &
\frac{g^2 N}{2 \pi^2\, \langle { \mathrm{Tr}} \, W(\cC) \rangle}
\langle {\mathrm{Tr}}\, \Phi_1(x_0,x_1) \hat F_{\mu \nu}(x_1)\cdot
\label{eq:two:probe:general} \\
& & \Phi_1(x_1,x_0)\Phi_2(x_0,x_2)
\hat F_{\alpha \beta}(x_2)\Phi_2(x_2,x_0) \, W(\cC_{x_0}) \rangle \,,
\nonumber
\end{eqnarray}
where the notations are the same as in Eq.~(\ref{eq:one:probe:general}). Graphically,
the correlator~(\ref{eq:two:probe:general}) is visualized in Figure~\ref{fig:probes}(bottom).

If the points $x_1$ and $x_2$ are approaching each other, $x_1 \to x$ and $x_2 \to x$,
and the orientations of the plaquettes coincide, $P_{\mu\nu}(x_1) = P_{\alpha\beta}(x_2)$,
then the internal parallel transporters cancel each other, $\Phi_1(x_1,x_0) \Phi_2(x_0,x_2) = \bbbone$,
and the quantum average~(\ref{eq:two:probe:general}) becomes similar to the expectation value
of the gluon condensate,
\begin{equation}
G_2(x) = \frac{g^2}{2 \pi^2} {\mathrm{Tr}}\, \hat F_{\mu\nu}(x) \hat F_{\mu\nu}(x)
\equiv
\frac{g^2}{4 \pi^2} F^a_{\mu\nu}(x) F^a_{\mu\nu}(x)\,,
\end{equation}
in the presence of the external meson.

In the exceptional case of the $SU(2)$ gauge group
\begin{equation}
\hat F_{\mu\nu}(x) \hat F_{\mu\nu}(x) = \frac{1}{4} F_{\mu\nu}^a(x) F^a_{\mu\nu}(x) \cdot \bbbone \equiv S(x) \cdot \bbbone\,,
\end{equation}
because in this case $T^a = \sigma^a/2$ ($\sigma^a$ are the Pauli matrices) and
$T^a T^b = \frac{1}{4}(\delta^{ab} + i \epsilon^{abc})$. Then the
probe~(\ref{eq:two:probe:general}) acquires a well--recognized meaning:
\begin{equation}
\cG_2^{\cC}(x) \equiv \cS^{\cC}_{\mu \nu \mu \nu }
= \frac{\langle G_2(x) \, {\mathrm{Tr}}\, W(\cC_{x_0}) \rangle}{\langle { \mathrm{Tr}} \, W(\cC) \rangle}
\,,
\label{eq:action:simplified}
\end{equation}
where the summation over the orientations of the common plaquette $P_{\mu\nu}(x)$ is implicitly assumed.
However in the physically relevant case of the $SU(3)$ gauge group one gets
$$T^a T^b = \frac{1}{6}\delta^{ab} + \frac{1}{2} d^{abc}+ i \frac{1}{2} \epsilon^{abc}\,,$$
where $f^{abc}$ is the totally asymmetric $SU(3)$ structure constant ($f^{abc} \equiv \epsilon^{abc}$
in the $SU(2)$ case), and $d^{abc}$ is the totally symmetric $SU(3)$ structure constant
($d^{abc} \equiv 0$ for the $SU(2)$ group). Thus, instead of the simple formula~(\ref{eq:action:simplified}),
in the $SU(3)$ case one gets the following relation:
\begin{eqnarray}
\cG_2^{\cC}(x) & = & \frac{1}{\langle { \mathrm{Tr}} \, W(\cC) \rangle}
\Bigl(\bigl\langle G_2(x)  {\mathrm{Tr}} W(\cC_{x_0}) \bigr\rangle \nonumber\\
& & + \frac{3 g^2}{4\pi^2} d^{abc} \bigl\langle F_{\mu\nu}^a(x) F^b_{\mu\nu}(x)
{\mathrm{Tr}} \, \Phi(x_0,x) T^c \Phi(x,x_0) W(\cC_{x_0}) \bigr\rangle\,\bigr)\,.
\nonumber
\end{eqnarray}
The first term fits our intuition about the correlation function between the gluon condensate and the flux tube, while
the second term makes the interpretation in terms of the gluon condensate difficult. Therefore, hereafter we consider
the $SU(2)$ case only (unless explicitly indicated otherwise). Our choice is also supported by the fact that
the correlations of the string with local observables are most studied in $SU(2)$ Yang--Mills theory.

Our next step is to evaluate the correlation function~(\ref{eq:two:probe:general}) in the field correlator approach.
Using the factorization of the correlator of the four field strengths operators~\cite{ref:susceptibility},
we get for the $N=2$ case:
\begin{equation}
\delta \cG^{\cC}_2(x) \equiv
\cG_2^{\cC}(x) - G_2 = - \frac{1}{3 \pi^2}
\cF^{\cC}_{\mu \nu}(x) \cF^{\cC}_{\mu \nu}(x)\,.
\label{eq:action:excess}
\end{equation}
Here $\cF^{\cC}_{\mu \nu}$ is the chromoelectric flux~(\ref{eq:one:probe:flux}) and
the v.e.v. of the condensate (i.e., the condensate calculated far from the string) is
\begin{equation}
G_2 = \frac{2}{\pi^2} D_{\mu \nu \mu \nu}(0) = \frac{12}{\pi^2} (D + D_1)(0)\,.
\label{eq:S:vac}
\end{equation}

\begin{figure}[htb]
\begin{center}
\vskip 5mm
\includegraphics[scale=0.6,clip=false]{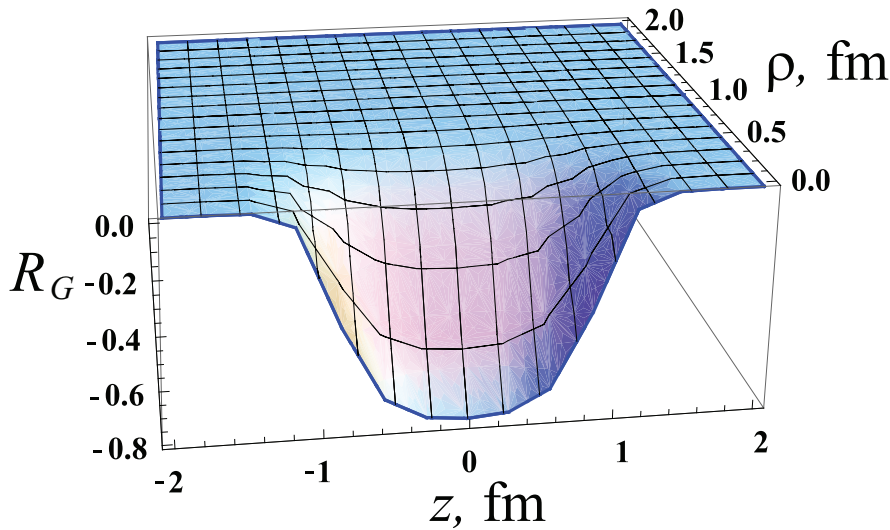}
\end{center}
\caption{The imprints of the string on the gluon condensate~(\ref{eq:RG}) and
on the topological charge density~(\ref{eq:Rchi}) calculated by the
method of field strength correlators. The transverse, $\rho$, and longitudinal, $z$,
coordinates with respect to the static string axis are shown in units of fermi.
The vertical axis is the excess of the gluon condensate~(\ref{eq:action:excess}) near the string
in units of the corresponding vacuum expectation value~(\ref{eq:S:vac}).
According to Eqs.~(\ref{eq:similarities}) and (\ref{eq:ratio}), the relative excess~(\ref{eq:Rchi}) of the topological
charge susceptibility~(\ref{eq:top:excess}) in units of the corresponding v.e.v.~(\ref{eq:chi:global:expr})
is twice larger in absolute value compared to the excess of the gluon condensate, $R_\chi = 2 \, R_G$.
The distance between the quark and the antiquark is 2~fm.}
\label{fig:profiles}
\end{figure}

The expression~(\ref{eq:action:excess}) represents the excess of the gluon condensate around the string.
Since the excess~(\ref{eq:action:excess}) is always negative we conclude that
the string suppresses the gluon condensate. This qualitative observation is in agreement with the
sum rules approach~\cite{ref:sum-rules}.
An example of the distribution of the gluon condensate around the string, expressed via the ratio
\begin{equation}
R_G(x) = \frac{\delta \cG^{\cC}_2(x)}{G_2}\,,
\label{eq:RG}
\end{equation}
is shown in Figure~\ref{fig:profiles}.
The shape of the distribution qualitatively resembles the results obtained in various numerical
simulations~\cite{ref:string:energy-action,Bissey,Chernodub:2005gz}.
A direct comparison of our analytical results and the results of the lattice simulations is unfortunately
not possible because in lattice simulations the string experiences the transverse quantum fluctuations contrary
to the case considered in this article.

\section{Topological charge fluctuations around the string}
The topological charge density is
\begin{equation}
q(x) = \frac{g^2}{16\pi^2} {\mathrm{Tr}}\, \hat F_{\alpha \beta}(x) \hat {\widetilde F}_{\alpha \beta}(x)\,,
\end{equation}
where $\hat {\widetilde F}_{\alpha \beta} = \epsilon_{\alpha \beta \mu \nu} \hat F_{\mu \nu}/2$.
Due to the CP--invariance of the Yang--Mills vacuum $\langle q(x) \rangle \equiv 0$. The presence of the string does
not violate that symmetry, and therefore $\langle q(x) \rangle_{\cC} \equiv 0$ as well.

The fluctuations of the topological charge in the four-dimensional volume $V$ are determined with
the help of the (global) topological susceptibility,
\begin{equation}
\chi_Q = \frac{1}{V} \langle (Q - \langle Q\rangle)^2 \rangle\,,
\qquad
Q = \int \dd^4 x\, q(x)\,,
\label{eq:chi:global}
\end{equation}
which was evaluated in with the help of the field correlator method in Ref.~\cite{ref:susceptibility}.
In this paper we are interested in local fluctuations of the topological charge, which are given
by the local susceptibility
\begin{equation}
\chi_q(x) = (q(x) - \langle q(x) \rangle)^2 \equiv q^2(x)\,.
\label{eq:chi:local}
\end{equation}
Here we used the subscript ``$q$'' in order to discriminate the local susceptibility~(\ref{eq:chi:local})
from the standard global one~(\ref{eq:chi:global}), $\chi_Q$.

In order to probe the distribution of the local fluctuations of the topological charge around the string
one needs at least four plaquette probes. The corresponding operator is constructed similarly to the
two-plaquette probe~(\ref{eq:two:probe:general}). Then we (i) chose the limit in which all four plaquettes are
approaching pairwise the two points $x_1$ and $x_2$, and (ii) sum properly over the orientations of the
corresponding plaquettes. The result is
\begin{equation}
\chi^{\cC} (x_1,x_2) = \frac{g^4}{64 \pi^4}
\frac{\langle {\mathrm{Tr}}\,
\hat \cF_{\mu \nu}(x_1) \, \hat {\widetilde \cF}_{\mu \nu}(x_1) \,
\hat \cF_{\alpha\beta}(x_2) \, \hat {\widetilde \cF}_{\alpha\beta}(x_2) \,
W(\cC_{x_0}) \rangle}{\langle { \mathrm{Tr}} \, W(\cC) \rangle}\,.
\label{eq:four:probe:general}
\end{equation}
In the $SU(2)$ gauge theory
\[
\hat F_{\mu \nu} \, \hat {\widetilde F}_{\mu \nu}
= \bbbone \cdot \frac{1}{4} F^a_{\mu \nu}\, {\widetilde F}^a_{\mu \nu}
= \bbbone \cdot \frac{1}{2} \Tr F_{\mu \nu}\, {\widetilde F}_{\mu \nu}\,,
\]
therefore
\begin{equation}
\chi^{\cC} (x) \equiv \chi^{\cC} (0,x) = \frac{\langle q(0) \, q(x) \, \Tr W(\cC) \rangle}{\langle \Tr W(\cC) \rangle}\,.
\label{eq:four:probe:su2}
\end{equation}

The quantity~(\ref{eq:four:probe:su2}) can be evaluated in the Gaussian approximation. In the leading order one gets
\begin{equation}
\chi^{\cC} (x) = \langle q(0) q(x) \rangle - \frac{D_{\alpha \beta\mu \nu}(0)}{144 \pi^4}
{\widetilde F}^\cC_{\alpha \beta}(0)
{\widetilde F}^\cC_{\mu \nu}(x)\,,
\end{equation}
where the first term in the right hand side is the topological charge correlator in the absence of the string.
This correlator was calculated in Ref.~\cite{ref:susceptibility}:
\[
\langle q(0) q(x) \rangle = \frac{(D+D_1)(x)}{4 \pi^4}
\Bigl[(D+D_1)(x) + x^2 {D_1}'\left(x^2\right)\Bigr]\,.
\]
As an interesting by-product we can also calculate the vacuum expectation value of the local
topological susceptibility given by the square of the topological charge density
in the $SU(N)$ gauge theory,
\begin{equation}
\chi_q \equiv \langle q^2(x) \rangle = \frac{3 N^2}{4(N^2-1)} \cdot \frac{1}{4 \pi^4} (D+D_1)^2(0)\,.
\label{eq:local:susceptibility}
\end{equation}
For a moment we restored the correct behaviour with respect to the number of colors $N$ by introducing the first prefactor in
the above equation. The local susceptibility~(\ref{eq:local:susceptibility}) looks differently from the global
susceptibility~(\ref{eq:chi:global}) which was calculated in Ref.~\cite{ref:susceptibility}:
\begin{equation}
\chi_Q = \frac{9 \, N^2 T^4_g}{64 \pi^2 (N^2 - 1)} D(0) (D + D_1)(0)\,.
\label{eq:chi:global:expr}
\end{equation}

In order to get the numerical value of the local susceptibility we use the data of Ref.~\cite{ref:new:fits}
in which the nonperturbative parts of Eqs.~(\ref{eq:D:D1}) were evaluated explicitly (the
superscript ``NP'' stands for ``nonper\-tur\-ba\-tive''):
\begin{equation}
D^{\NP}(0) = 0.212(11)\,\mbox{GeV}^4\,,
\qquad
D^{\NP}_1(0) = 0.072(4) \,\mbox{GeV}^4\,,
\qquad
T_g = 0.222(4)\,\mbox{fm}\,.
\end{equation}
Substituting these numerical values into Eq.~(\ref{eq:local:susceptibility}) we get the following result for
the local topological susceptibility in $SU(3)$ Yang-Mills theory:
\begin{equation}
\chi_q \equiv \langle q^2(x) \rangle = {[340(4) \, \mbox{MeV}]}^8\,.
\label{eq:chi:loc:value}
\end{equation}
This value can be compared with the global susceptibility~(\ref{eq:chi:global:expr}),
Ref.~\cite{ref:susceptibility}:
\begin{equation}
\chi_Q = {[196(7) \, \mbox{MeV}]}^4\,.
\end{equation}
Both quantities have the same magnitude in the energy scale.

We are interested in the imprint of the string on the topological charge fluctuations characterized
by the excess,
\begin{equation}
\delta \chi^{\cC} (x) = \chi^{\cC} (x) - \chi_q\,,
\end{equation}
of the value of the local
susceptibility~(\ref{eq:chi:local}) near the string with respect
to the vacuum expectation value~(\ref{eq:local:susceptibility}), (\ref{eq:chi:loc:value}):
\begin{equation}
\delta \chi^{\cC} (x) =
- \frac{(D+D_1)(0)}{72 \pi^4} \cF^\cC_{\mu \nu}(0) \cF^\cC_{\mu \nu}(x)\,.
\label{eq:top:excess}
\end{equation}

Equation~(\ref{eq:top:excess}) demonstrates clearly that the fluctuations of the topological charge density in
the vicinity of the string are smaller compared to the fluctuations far outside the string. Moreover, the string
affects the fluctuations of the topological charge~(\ref{eq:top:excess}) essentially in the same way as it acts on the
gluon condensate~(\ref{eq:action:excess}). The illustration of this effect can be characterized by the ratio
\begin{equation}
R_\chi(x) = \frac{\delta \chi^{\cC} (x)}{\chi_Q}\,.
\label{eq:Rchi}
\end{equation}
This quantity is shown in Figure~\ref{fig:profiles}.

It is possible to relate the shape of the string in terms of the topological susceptibility~(\ref{eq:top:excess})
with the imprint of the string on the gluon condensate~(\ref{eq:action:excess}):
\begin{equation}
\delta \chi^{\cC} (x) = \frac{G_2}{288} \delta \cG_2^{\cC}(x)\,.
\label{eq:similarities}
\end{equation}
Thus, in the leading order the shapes of the string, imprinted on the topological charge density and
in the gluon condensate are proportional to each other. In particular, the transverse width of the string
in both variables should essentially be the same, $\lambda = T_g/2$.

It is also interesting to check the strength of the suppression on the axis of the infinitely long string. One finds for
the gluon condensate and topological charge density, respectively,
\begin{eqnarray}
\delta \chi^{\cC} (0) & = & - \frac{4 \, T_g^4}{9 \pi^2} D^2(0) (D+D_1)(0) = - (415\,\mbox{MeV})^8\,,
\nonumber\\
\delta \cG_2^{\cC} (0) & = & - \frac{32 T_g^4}{3} D^2(0) = - (0.92\,\mbox{GeV})^4\,.
\nonumber
\end{eqnarray}

The suppression of the condensate and the susceptibility at the axis of the infinitely long string can be compared with
the corresponding v.e.v.'s far from the string [given by Eq.~(\ref{eq:S:vac}) and Eq.~(\ref{eq:chi:global:expr}), respectively].
The result is
\begin{equation}
|R_G(0)| = \frac{1}{2} |R_\chi(0)|
= \frac{8 \pi^2}{9} \frac{T^4_g D^2(0)}{(D + D_1)(0)} \approx 2.1\,,
\label{eq:ratio}
\end{equation}
where the ratios $R_G$ and $R_\chi$ are presented in Eqs.~(\ref{eq:RG}) and (\ref{eq:Rchi}), respectively.
It turns out that the suppression for the non-fluctuating string is very strong: at the string axis
the gluon condensate (the topological susceptibility) become negative in sign and twice (four times)
larger in the absolute value compared to the value far from the string.

\section{Conclusions}
We studied the structure of the confining string in terms of the gluon condensate
and the topological charge susceptibility (the topological charge density squared).
The calculations -- performed in $SU(2)$ Yang-Mills theory with the help of the method of
the field strength correlators -- show that the both quantities are suppressed in the vicinity of
the string axis. Qualitatively, we found the agreement with the corresponding results of the lattice simulations.
Quantitatively, it is hard to compare the level of the on-axis suppression with the analogous results
of the lattice numerical simulations because of the string widening effect in the latter case.
We also have found the vacuum expectation value of the topological charge density
squared~(\ref{eq:local:susceptibility}). The numerical value in the case of the
$SU(3)$ gauge group is given in Eq.~(\ref{eq:chi:loc:value}).

\section*{Acknowledgments}
M.N.Ch. is thankful to the members of Laboratoire de Mathematiques et Physique Theorique
of Tours University for hospitality and stimulating environment. The work is supported by Federal Program of
the Russian Ministry of Industry, Science and Technology No. 40.052.1.1.1112, by the grants RFBR 05-02-16206a,
RFBR-DFG 06-02-04010, by a STINT Institutional grant IG2004-2 025 and by a CNRS grant.

\end{document}